# On the electron sheath theory and its applications in plasma-surface interaction

Guang-Yu Sun[*], Zhang Shu, An-Bang Sun[†], and Guan-Jun Zhang[‡]

State Key Laboratory of Electrical Insulation and Power Equipment, Xi'an Jiaotong University, School of Electrical Engineering, Xi'an, Shaanxi, 710049, China

The electron sheath is a particular electron-rich sheath with negative net charges where plasma potential is lower than the biased electrode. Here an improved understanding of electron sheath theory is provided using both fluid and kinetic approaches while elaborating on its implications for plasma-surface interaction. A fluid model is first proposed considering the electron presheath structure, avoiding the singularity in electron sheath Child-Langmuir law. The latter is proved to underestimate the sheath potential. Subsequently, the kinetic model of electron sheath is established, showing considerably different sheath profiles in respect to the fluid model due to the electron velocity distribution function and finite ion temperature. The model is then further generalized involving a more realistic truncated ion velocity distribution function. It is demonstrated that such distribution function yields a super-thermal electron sheath whose entering velocity at sheath edge is greater than that prescribed by the Bohm criterion, implying a potentially omitted calibration issue in the probe measurement. Furthermore, an attempt is made to incorporate the self-consistent presheath-sheath match within the kinetic framework, showing a necessary compromise between realistic sheath entrance and the inclusion of kinetic effects. In the end, the consequent secondary electron emission due to sheath-accelerated plasma electrons in electron sheath are analyzed, providing a sheath potential coupled with the plasma and wall properties.

## 1 Introduction

The sheath is a space charge region commonly formed in the edge of plasma which breaks up the quasi-neutrality and shields the bulk plasma from solid boundary. Sheath plays essential role in confined plasma research in terms of plasma-surface interaction: plasma processing, plasma propulsion engine, magnetic-confined fusion, dust particles, plasma diagnostics, etc.[1-6] A classic Debye sheath (aka ion sheath) appears when a floating slab is injected into a plasma. Initially, more electrons than ions enter the board due to higher mobility, leaving net positive charges in the gradually formed "sheath" region while depositing negative charges in the solid material, which in turn mitigates the electron flow until plasma current is balanced at the solid wall. Though such kind of ion-rich sheath is most frequently encountered, a sheath can be electron-rich in some particular cases. One example is when an electrode is biased above plasma potential, where an electron sheath instead of ion sheath is formed near the electrode.[7]

[*] Current address: Ecole Polytechnique Fédérale de Lausanne (EPFL), Swiss Plasma Center, Lausanne, CH-1015, Switzerland.
[†] Email: anbang.sun@xjtu.edu.cn
[‡] Email: gjzhang@xjtu.edu.cn

The electron sheath is featured by the increasing potential from the sheath edge to the solid boundary, thus plasma electrons are accelerated by sheath potential whereas ions are deaccelerated, contrary to the ion sheath. The electron sheath is a relatively local phenomenon and requires concomitant ion sheath to present in other plasma-facing components to achieve the global particle balance in a confined plasma system.[8] If the surface areas where electron and ion sheath presents are noted as $A_w$ and $A_c$, with subscript $w$ and $c$ representing the electrode and other chamber wall, then the total electron current to all surfaces becomes $I_{e,tot} = e\Gamma_{ep}[A_w + A_c \exp(-\frac{e\varphi_{pc}}{T_{ep}})]$, with $\Gamma_{ep}$ the electron flux at sheath edge and $\varphi_{pc}$ the potential difference between other chamber wall and plasma (not between the electrode and plasma) which is positive. Total ion flux is $J_{i,tot} = e\Gamma_i A_c$, with $\Gamma_i$ the ion flux at sheath edge. Note that ion flux reaching the electrode where electron sheath presents is neglected due to limited ion temperature. Assuming Maxwellian electron and ion sheath Bohm criterion, equating the ion and electron total current gives:[9]

$$e\varphi_{pc} = -T_{ep}\ln(\sqrt{\frac{2\pi m_e}{m_i}} - \frac{A_w}{A_c}) \quad (1)$$

Here $m_i$ and $m_e$ are ion and electron mass, respectively. The equation is the same as floating sheath potential expression apart from the $-\frac{A_w}{A_c}$ term. Equation (1) also requires that $A_w \leq A_c\sqrt{\frac{2\pi m_e}{m_i}}$ which stipulates a very small surface area where electron sheath presents.

Apart from the Langmuir probe sweep in the electron saturation region, electron sheath also appears in a variety of occasions in dusty plasma particle circulation, scrape-off layer diagnostics, planet surface, spacecraft probe, anode ablation by arc, and some transient processes,[10-14]. Nonetheless, the related studies on electron sheath are far less than those of the ion sheath, and the electron sheath counterparts of some well-developed modeling in ion sheath have yet to be established.

Classic understanding of electron sheath is primarily based on the pragmatic demand for probe diagnostics. For simplicity, a probe biased more positively than the plasma is assumed to collect the full electron current from the electron sheath, while the presence of ion in sheath is neglected. The electron velocity distribution function (EVDF) of sheath entrance is trivially regarded as half-Maxwellian and no presheath structure is considered.[15] Such assumption greatly facilitates the implementation of electron sheath physics into probe calibration. However, recent progresses of electron sheath theories showed that the underlying physics are not that simple. Simulation of Yee et al proved that a presheath exists for electron sheath which accelerates the electron up to the thermal velocity in sheath entrance.[16] The presheath size for electron sheath is longer than ion presheath and extends deep into the bulk plasma. Also, a comprehensive presheath theory was proposed by Brett et al based on fluid equations.[17] It was shown that electrons in presheath are mainly driven by pressure gradient, and that the accelerated electrons give rise to fluctuations due to ion acoustic instabilities. Instabilities in electron sheath were also investigated experimentally by Stenzel et al.[18, 19] Additionally, the transition between electron sheath and ion sheath was reproduced in simulation by varying electrode biased potential.[20] The electron sheath can even be enclosed by an ion presheath by surrounding a metallic electrode with dielectric material. [21]

So far, the existing electron sheath theories are mostly based on fluid model and mainly focus on the presheath region while the sheath is only treated as a boundary condition. The fluid model is appropriate in the presheath region as collisionality there is sufficiently large to form a Maxwellian distribution. However, the stories in the electron sheath region are not so simple as we will demonstrate below. For one thing, the newly discovered electron entering velocity at sheath edge inevitably yields a truncated electron velocity distribution function at sheath entrance. The ion entering velocity in ion sheath edge barely matters since initial ion energy is far lower than the ion sheath potential energy, whereas this is not the case in electron sheath. For another, the plasma electrons are accelerated by the sheath potential and collide on the wall with greater energy compared with the electrons in ion sheath, inducing secondary electron emission (SEE). This may alter the current balance since both emitted electrons and plasma electrons contribute to the measured current, yet such effect has drawn few attentions. The present work attempts to address these issues on a theoretical ground and discuss the pertinent implications in plasma-surface interactions.

The paper structure is given as follows. In section 2 the fluid model of electron sheath is constructed and is compared with the Child-Langmuir law prediction. Section 3 provides a comprehensive kinetic modeling of electron sheath, involving electron and ion velocity distribution function and the influence of electron-ion temperature ratio. In section 4 the secondary electron emission caused by sheath accelerated electrons is incorporated into the electron sheath model. Concluding remarks are given in section 5.

## 2 A fluid view of the electron sheath

### 2.1 Revisit of Child-Langmuir description in an electron sheath

Child-Langmuir law is widely used for the descriptions of high voltage sheath, i.e. plasma-facing electrode is strongly biased so that the sheath potential is much larger than electron temperature as well as floating sheath potential. For high voltage ion sheath, the sheath potential is high enough so that plasma electrons can hardly penetrate the sheath, therefore only ion density is counted in sheath. Different from the matrix sheath model where a uniform ion density in sheath is assumed, the Child-Langmuir model considers the acceleration of ions in sheath and the consequent decreasing ion density towards the boundary. When the electrode is biased increasingly negative with respect to the plasma, a transition from floating ion sheath to Child law sheath occurs. The ion sheath potential of a floating, non-emissive boundary can be easily obtained from the balance of electron and ion fluxes. Here electron flux is $\Gamma_{ep} = n_{se}\sqrt{\frac{T_{ep}}{2\pi m_e}}\exp(\frac{e\varphi_{ish}}{T_{ep}})$ , with $\varphi_{ish}$ the ion sheath potential and $T_{ep}$ the plasma electron temperature, and ion flux is $\Gamma_i = n_{se}\sqrt{\frac{T_{ep}}{m_i}}$. The current balance gives:

$$e\varphi_{ish} = -\frac{T_{ep}}{2}\ln(\frac{\mu}{2\pi}) \quad (2)$$

with $\mu = m_i/m_e$, $n_{se}$ the plasma density at sheath edge (quasi-neutrality is assumed), and ion sheath potential $\varphi_{ish} < 0$, note that by convention the potential at sheath edge is assumed 0. When rising up the biased voltage (absolute value), the flux balance at electrode is no longer valid and a net current term $J_{net}$ must appear to compensate: $\Gamma_i - \Gamma_{ep} = \Gamma_{net} = J_{net}/e$. In the high voltage sheath limit, electron flux term is dropped and the Child-Langmuir law for ion sheath as we as potential distribution is derived as:

$$J_{net} = \frac{4\varepsilon_0}{9x_{ish}^2}\sqrt{\frac{2e}{m_i}}(-\varphi_{ish})^{3/2} \quad (3)$$

$$\frac{x}{\lambda_{De}} = 0.79(\frac{-e\varphi_{ish}}{T_{ep}})^{3/4} \quad (4)$$

where $x_{ish}$ is ion sheath length and $\lambda_{De}$ is electron Debye length. Detailed derivations are available in numerous references and are not to be repeated here.[22] In the case of an electron sheath, however, the above floating to high voltage sheath transition is somewhat different. In an electron sheath the potential relative to sheath edge is positive, electrons are accelerated while ions are deaccelerated. Obviously if one applies similar method as in the ion sheath, the floating electron sheath potential satisfies $n_{se}\sqrt{\frac{T_{ep}}{2\pi m_e}} = n_{se}\sqrt{\frac{T_i}{2\pi m_i}}\exp\left(-\frac{e\varphi_{esh}}{T_i}\right)$ with LHS the electron flux and RHS the ion flux, if ion velocity distribution function (IVDF) is assumed Maxwellian. Here $\varphi_{esh}$ is electron sheath potential (again relative to sheath edge) and $T_i$ is ion temperature. This gives $e\varphi_{esh} = \frac{T_i}{2}\ln\left(\frac{T_i}{\mu T_{ep}}\right) < 0$ in typical plasma conditions, contradicting to our assumption of positive electron sheath potential. This is because the ion and electron flux cannot be balanced solely on one electrode for the electron sheath. As discussed above, one criterion for the existence of electron sheath is that the corresponding surface area where electron sheath presents should be smaller than ion sheath surface area in a discharge chamber.

The high voltage sheath limit of electron is similar to the ion sheath. Ignoring initial electron energy at sheath and assuming no source term in the continuity equation, the plasma electron density is simply:

$$n_{ep}(\varphi) = \frac{J_{net}}{e}(\frac{2e\varphi}{m_e})^{-0.5} \tag{5}$$

Note that the electrons velocity distribution function is not considered here. Ignoring ion density, the Poisson equation is written as:

$$\frac{d^2\varphi}{dx^2} = \frac{J}{\varepsilon_0}(\frac{2e\varphi}{m_i})^{-0.5} \tag{6}$$

with $\varepsilon_0$ the vacuum permittivity. Multiply both sides in Equation (6) by $\frac{d\varphi}{dx}$ and intergrate from sheath edge to position $x$ with respect to $dx$ in the electron sheath, we arrive at:

$$\frac{d\varphi}{dx} = \sqrt{\frac{4J_{net}}{\varepsilon_0}(\frac{m_i}{2e})^{0.5}\varphi^{0.25}} \tag{7}$$

Note that the potential and electric field are assumed 0 at sheath edge in deriving Equation (7). Integrating above equation again and normalizing the $x$ coordinate with regard to electron Debye length $\lambda_{De}$, the Child-Langmuir law of electron sheath and potential distribution are derived as follows:

$$J_{net} = \frac{4\varepsilon_0}{9x_{esh}^2}\sqrt{\frac{2e}{m_e}}(\varphi_{esh})^{3/2} \tag{8}$$

$$\frac{x}{\lambda_{De}} = 1.26(\frac{e\varphi_{esh}}{T_{ep}})^{3/4} \tag{9}$$

One can see that the relations between sheath potential and current are similar in both electron and ion sheath. While comparing Equation (4) and (9), the size of electron sheath is slightly larger than that of the ion sheath if the absolute values of electron and ion sheath potential are equal. This feature has been verified experimentally by quantifying space potential in sheath using emissive probe.[23]

An important assumption on the deduction above is the omission of initial electron energy at sheath edge. The ion velocity at sheath edge, in high voltage ion sheath, is commonly negligible when deriving the Child-Langmuir law since the sheath potential is large with respect to the electron temperature. As discussed in Section 1, the electron presheath is not considered by the conventional description. It is only recently that special attention is drawn regarding the importance of presheath in electron sheath properties.[17, 24] Electron entering velocity at sheath edge was shown to be $\sqrt{m_i/m_e}$ times larger than ion Bohm velocity. Meanwhile, the electrode biased voltage is usually smaller in electron sheath compared with ion sheath in typical plasma applications. The treatment of electron entering velocity should therefore be reviewed in section 2.2, where it will be shown that such revision actually brings remarkable influence in the obtained sheath properties.

Another issue of the above deductions is the neglected electron velocity distribution function. Equation (4) implicitly regards electrons as monoenergetic beam. Though this is also used in the derivation of ion sheath Child-Langmuir law, one has to keep in mind that the electron temperature is far greater than ion temperature for a typical cold plasma discharge, thus the omission of kinetic effects will introduce larger discrepancies compared with that in ion sheath. Further discussion on the treatment of EVDF will be given in section 3.

### 2.2 Fluid description of electron sheath with presheath

To include the influence of electron entering velocity, it is intuitive to write down the electron fluid equations in sheath and regard the entering velocity as a boundary condition. To begin with, the particle and momentum balance of electron in sheath is given as follows:

$$\frac{d}{dx}\left(n_{ep}u_e\right) = 0 \tag{10}$$

$$\frac{d}{dx}\left(m_e n_{ep} u_e^2 + P_{e\parallel}\right) = -e n_{ep} E \tag{11}$$

In above equation $u_e$ is electron fluid velocity, $P_{e\parallel}$ is parallel electron pressure, and $E$ is electric field. Source terms are neglected since the sheath is assumed collisionless. Combining equation (10), (11) and solving for $u_e$ with potential $\varphi$, the following relation is obtained:

$$\int_{u_{eo}}^{u_e}(u'_e - \frac{T_{ep}}{m_e u'_e})du'_e = \frac{e}{m_e}\int_0^{\varphi} d\varphi' \tag{12}$$

where $u_{eo}$ is the electron fluid velocity at sheath edge and is treated as the boundary condition, also the potential at sheath edge is chosen as 0. Note that here the temperature gradient is dropped and the isothermal relation is used. The resulting equation is:

$$\frac{1}{2}(\frac{u_e}{u_{e0}})^2 - \ln\left(\frac{u_e}{u_{e0}}\right) - 0.5 = \frac{e\varphi}{T_{ep}} \tag{13}$$

Equation (13) can be solved by rewriting it in the form of Lambert W function and taking the asymptotic limit at large potential. More details on the solution of this form of equation were given in the in the work of Brett et al[17], here we give the result directly:

$$u_e = u_{e0}\sqrt{1 + 2e\varphi/T_{ep}} \tag{14}$$

It is then possible to solve the Poisson equation in the following form:

$$\frac{d^2\Phi}{dX^2} = (1 + 2\Phi)^{-0.5} \tag{15}$$

Here the electron continuity equation ($n_{se}u_{e0} = n_e u_e$) is used again and the following normalized terms are adopted for simplicity.

$$\Phi = \frac{e\varphi}{T_{ep}}, X = \frac{x}{\lambda_{De}} \tag{16}$$

The potential distribution in electron sheath is then obtained by multiplying Equation (15) by $\frac{d\Phi}{dX}$ and integrating twice with respect to $X$, which is reduced to:

$$X = \frac{\sqrt{2}}{6}(2\sqrt{2\Phi + 1} + 4)\sqrt{(2\Phi + 1)^{0.5} - 1} \tag{17}$$

The potential distributions given by Equation (9) and (17) are shown in Figure 1 (named separately as the Child-Langmuir model and the revised fluid model). The difference is minor in low potential range but becomes remarkable at high potential levels. Generally, the Child-Langmuir solution slightly overestimates the electron sheath potential in low potential range but strongly underestimates the potential at high potential levels, at given electron sheath size. One may also conclude that the Child-Langmuir solution predicts smaller sheath size for low biased potential but larger sheath size for high biased potential.

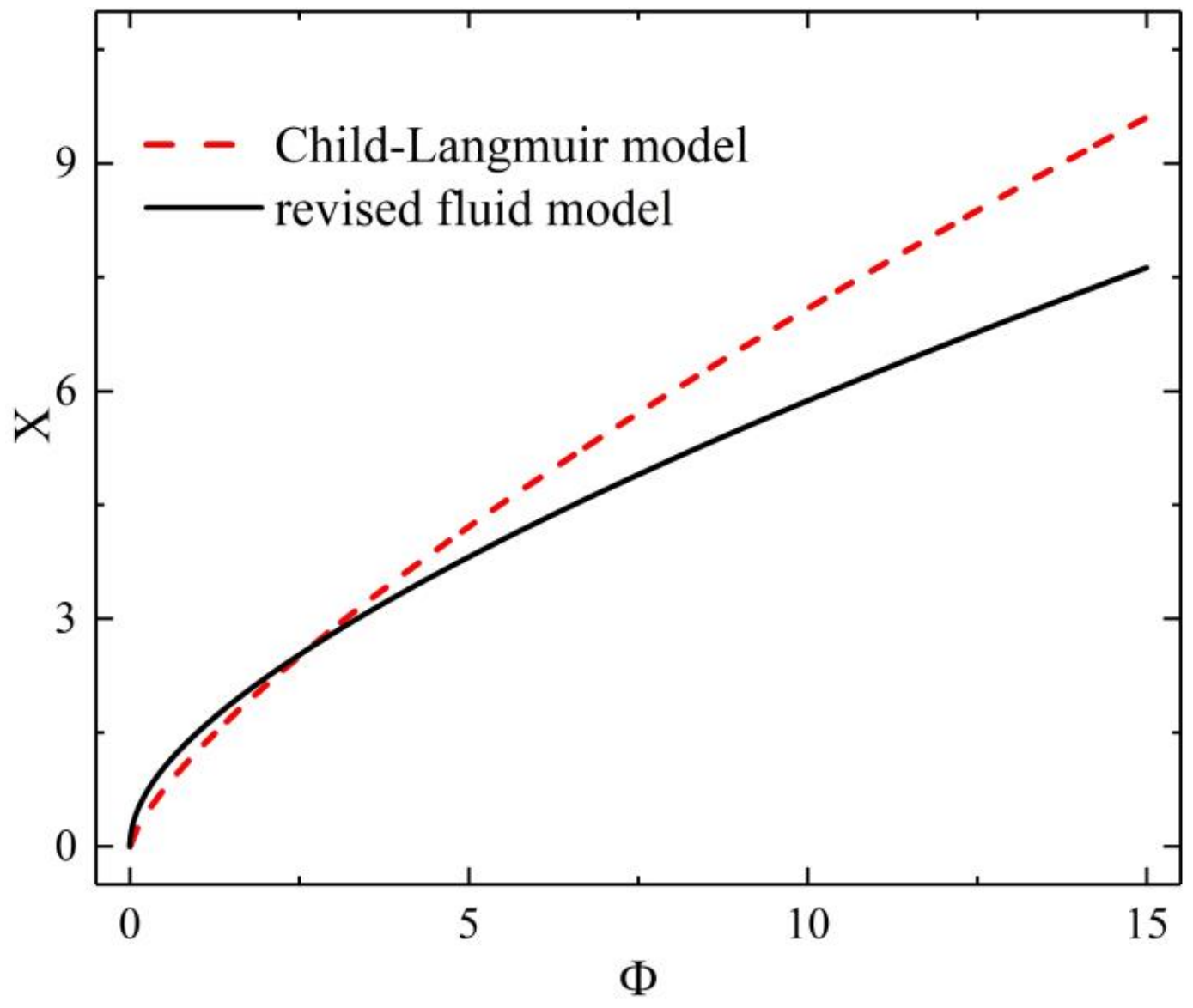

Figure 1. Comparison of potential distribution in electron sheath based on Child-Langmuir law and revised fluid model. $\Phi$ is normalized potential and $X$ is normalized position. At given sheath size, the Child-Langmuir model overestimates the sheath potential in low potential range and underestimates the sheath potential at high potential levels.

The electron density distributions in sheath predicted by Child-Langmuir model and revised fluid model are shown in Figure 2. To make the comparison, the current term $J$ in Equation (5) is replaced by the half-Maxwellian electron flux expression at sheath edge times the elementary charge, assuming the full electron current is collected at the electrode. The Child-Langmuir model generally underestimates the electron density in locations far away from the sheath edge. One major flaw of Child-Langmuir model is that there exists a density singularity at $X = 0$ due to the neglection of electron entering velocity at sheath edge. We will show in section 3.1 that the singularity can also be avoided by considering electron kinetic effects even without involving the entering velocity.

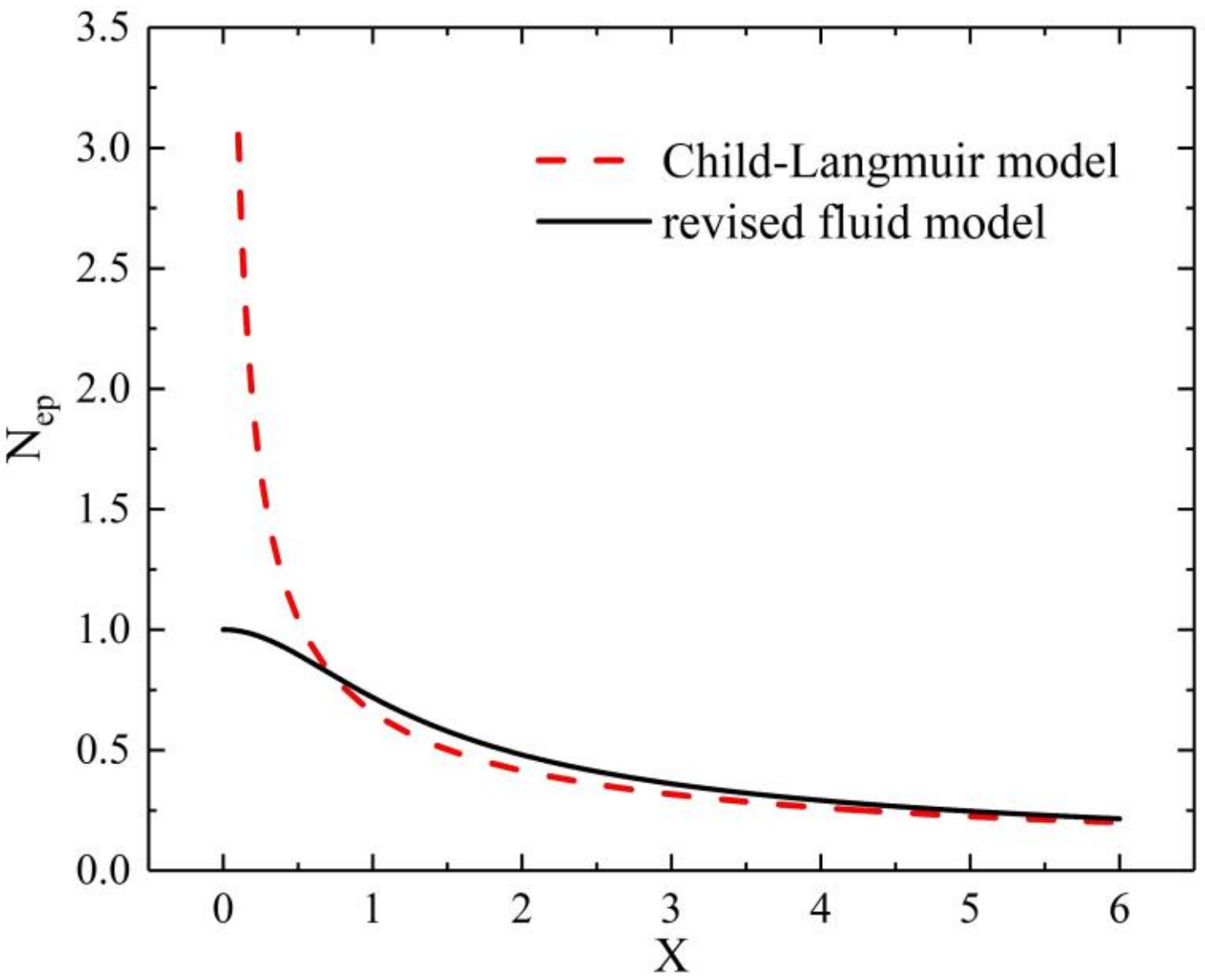

Figure 2. Normalized electron density distribution in sheath region predicted by Child-Langmuir model and revised fluid model. Child-Langmuir model gives infinite density at $X = 0$. The part of Child-Langmuir solution curve in the vicinity of singularity is not shown for better figure scaling. The Child-Langmuir law predicts lower electron density except near $X = 0$.

It is also important to point out that the obtained conclusions from Equation (17) seem to be independent from the exact value of the entering velocity. This doesn't mean that the value of $u_{e0}$ is irrelevant, rather it is just because the electron sheath cannot achieve self-consistency isolated from the presheath. Firstly, the $u_{e0} = 0$ assumption adopted in Child-Langmuir solution will lead to unphysical solution of $u_e = 0$ everywhere according to Equation (14). The $u_{e0} = 0$ assumption also leads to a singularity at sheath edge in Equation (5). Hence the inclusion of electron entering velocity is crucial. In addition, the choice of $u_{e0}$ is not arbitrary, which is rather dictated by the way the presheath matches the sheath. We will show in section 3.2 that the entering velocity can always be expressed as $u_{e0} = \alpha_{ue0}\sqrt{\frac{T_{ep}}{m_e}}$ with $\alpha_{ue0}$ a coefficient depending on the electron and ion model (fluid or kinetic), their temperatures, and choice of distribution function if kinetic model is employed. But at given model and plasma parameters, the value of $u_{e0}$ is definite. This subject will be further developed in section 3.2.

# 3 A kinetic view of the electron sheath

## 3.1 Electron sheath solution using kinetic model

In section 2, the electron fluid equations are used to derive the electron sheath solution. However, the exact EVDF inside the sheath is not considered. To illustrate the influence of kinetic effects on the electron sheath solution, here we attempt to establish the electron sheath theory in a kinetic approach. In addition, ion distribution is involved to investigate the influence of ion temperature for more general applications of the theory.

To begin with, the electron and ion densities inside the sheath should be determined. The electrons are accelerated due to increasing sheath potential towards the electrode, hence a velocity lower bound appears when integrating the EVDF which gives the following electron density:

$$n_{ep} = n_{se}\exp(\frac{e\varphi}{T_{ep}})\mathrm{erfc}(\sqrt{\frac{e\varphi}{T_{ep}}}) \tag{18}$$

The ions are repelled by electron sheath potential just as the electrons in an ion sheath. If the force posed by electron field is balanced by the pressure gradient force, the Boltzmann distribution could be used to give the following ion density:

$$n_i = n_{se}\exp(-\frac{e\varphi}{T_{ei}}) \tag{19}$$

The Poisson equation can therefore be written as:

$$\frac{d^2\Phi}{dX^2} = \exp(\Phi)\,\mathrm{erfc}(\sqrt{\Phi}) - \exp(-\Theta_T\Phi) \tag{20}$$

with $\Theta_T = T_{ep}/T_i$ the ratio of electron and ion temperature. Different from Equation (6) and (15), Equation (20) cannot be solved fully analytically. A possible solution is to reduce Equation (20) into the following form, then solve it as an initial value problem (IVP) numerically:

$$\frac{d\Phi}{dX} = \sqrt{2}[\exp(\Phi)\,\mathrm{erfc}(\sqrt{\Phi}) - 1 + 2\sqrt{\frac{\Phi}{\pi}} + \exp(\Theta_T\Phi)]^{0.5} \tag{21}$$

The equation above is obtained again by multiplying Equation (20) by $\frac{d\Phi}{dX}$ and integrating twice with respect to $dX$. To solve for the potential numerically, the electrode potential relative to sheath edge is given as the initial condition. The potential is then solved towards the sheath edge using numerical methods. Here the explicit Euler method is employed:

$$\Phi_{n+1} = \Phi_n + g(\Phi_n)\Delta X, n = 1,2,3 \dots \quad (22)$$

with $g(\Phi)$ the RHS of Equation (21), $\Delta X$ the position step size. The initial condition at electrode is the normalized electron sheath potential: $\Phi_1 = \Phi_w = \Phi_{esh}$. $\Phi_w$ is the normalized potential at electrode and will constantly appear in the following derivations.

Calculated results with different $\Phi_w$ values are shown in Figure 3(a) and are compared with the Child-Langmuir predictions. One can found that the sheath size given by the kinetic model is actually not far from that given by the Child-Langmuir solution, but the profile of sheath potential distribution in space are remarkably different. In general, kinetic model predicts a shorter sheath size at fixed electrode potential, and its space potential profile is higher than the Child-Langmuir solution even the applied electrode potential is the same.

The influence of ion temperature is shown in Figure 3(b). It is clear that the ion temperature only exerts minor influence on the shape of potential profile, and high ion temperature slightly decreases the size of electron sheath. The potential profile becomes virtually insensitive to ion temperature after the value of $\Theta_T$ exceeds around 5. Consequently, in typical low temperature plasma applications, the influence of ion temperature on potential distribution is marginal. The calculated electron and ion density distributions in sheath predicted by kinetic model are compared with Child-Langmuir model in Figure 4. The electron and ion densities are normalized with respect to the sheath edge density:

$$N = \frac{n}{n_{se}} \quad (23)$$

It can be seen that generally the kinetic model gives lower electron density compared with Child-Langmuir model, suggesting an overestimation of plasma density when using fluid model. The kinetic model avoids the singularity at $X = 0$ though both the kinetic model here and the Child-Langmuir model disregard the electron presheath structure. The electron density profile is barely touched by the change of $\Theta_T$, whereas the ion density profile changes remarkably with $\Theta_T$. Note that here we adopt the ion Boltzmann distribution, whereas in practice, some energetic ions might penetrate the electron sheath and arrive at electrodes, which is increasingly obvious for low $\Theta_T$. These ions will not return to plasma and inevitably leaves an IVDF dissipated at high velocity tail.[25] Also, the ion density could be slightly higher than electron density in the vicinity of sheath edge. This actually contradicts the general Bohm criterion, and is due to the neglection of electron presheath structure in the kinetic model. Further analyses regarding more accurate IVDF and the influence of electron presheath structure on sheath solution will be given in section 3.2 and 3.3.

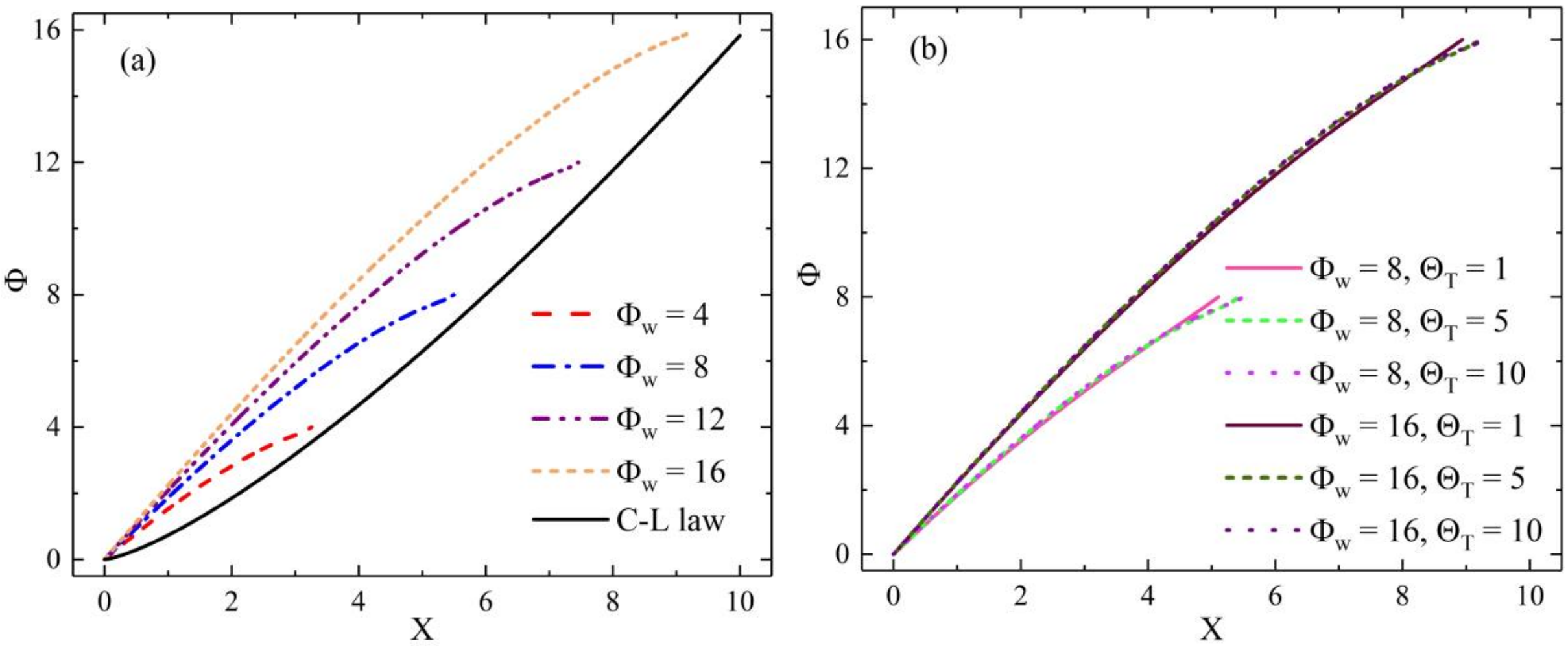

Figure 3. (a) The comparison of sheath potential distributions given by adopted kinetic model and Child-Langmuir law. $X$ is normalized position and $\Phi$ is normalized potential. (b) The influence of Electron-ion temperature ratio on sheath potential distribution. The influence is negligible after $\Theta_T$ surpasses 5.

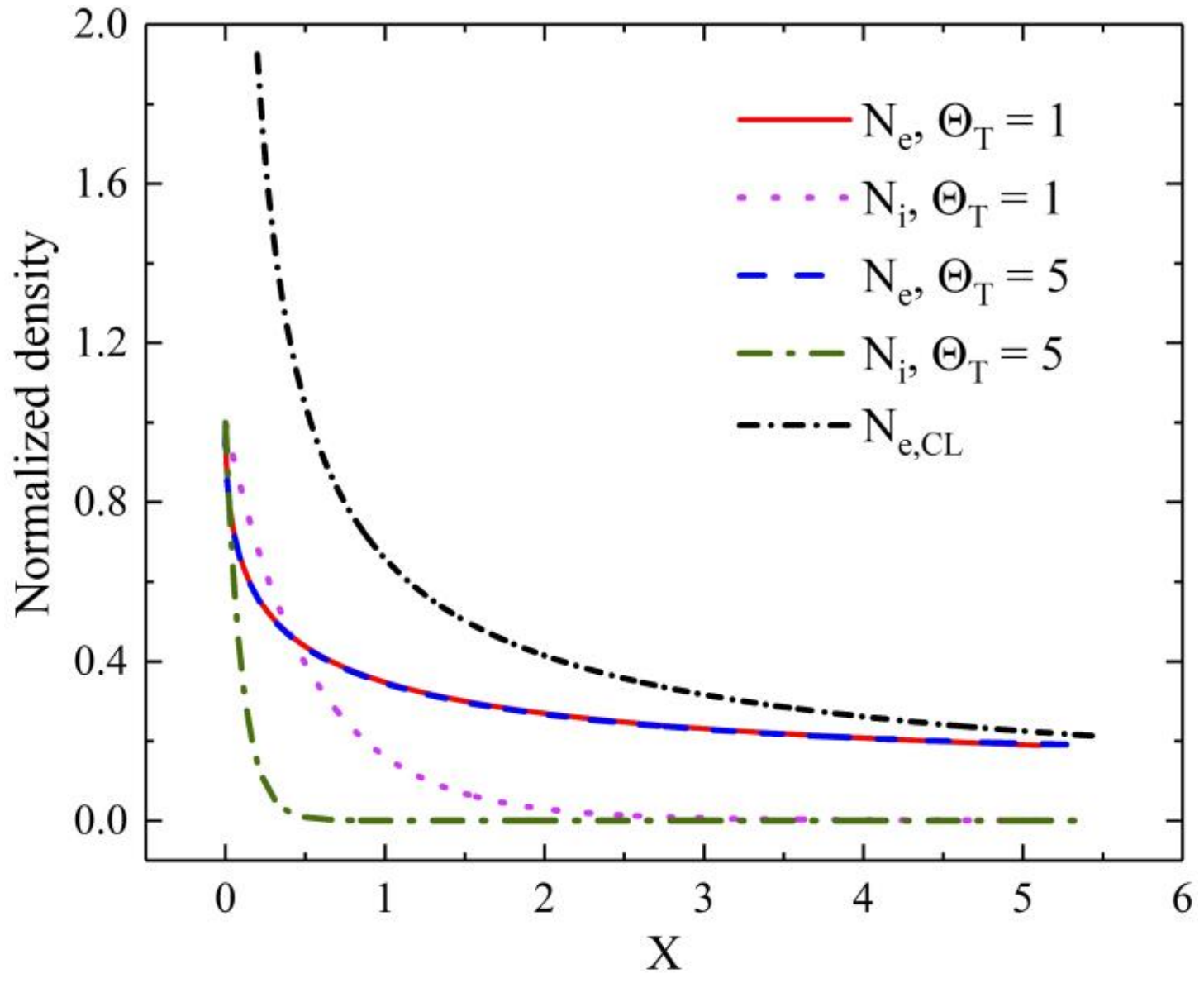

Figure 4. The electron and ion densities given by kinetic model and Child-Langmuir model. Different electron-ion temperature ratios are used to illustrate influence of ion temperature. The part of Child-

Langmuir solution curve in the vicinity of the singularity is not shown for better figure scaling. Electron density profile is not sensitive to $\Theta_T$ while ion density profiles change remarkably with $\Theta_T$.

## 3.2 Improved modeling of electron sheath Bohm criterion

In section 2, we have illustrated the influence of electron entering velocity on electron sheath solution, where the inclusion of electron fluid velocity $u_{e0}$ as a boundary condition in sheath entrance shifts the size and potential distribution of the electron sheath. Judging solely from the sheath side, however, the value of $u_{e0}$ seems arbitrary. In this section, the electron sheath will be patched to the presheath which dictates the choice of $u_{e0}$. We will first review the general Bohm criterion and apply it specifically in the electron sheath, then the modified Bohm criterion will be proposed considering more realistic IVDF. The obtained conclusions will be implemented into the following kinetic modeling of the electron sheath in various conditions.

### 3.2.1 The general Bohm criterion and derivation of electron sheath Bohm criterion

The common procedure of the derivation of Bohm criterion for electron sheath is to calculate the electron and ion density at sheath edge and apply the general Bohm criterion, which is more universal and applies to both electron and ion sheath. The general Bohm criterion is frequently used directly and was originally proposed in Riemann's work.[26] However, sometimes its exact form could be vague in literatures and may cause misunderstanding when applied to ion and electron sheath. Here the general Bohm criterion is briefly reviewed and then directly applied to the electron sheath.

We start from the Poisson equation and expand it at sheath edge:

$$\nabla^2\varphi = \frac{-1}{\varepsilon_0}\left[\rho|_{\varphi=0} + \left.\frac{d\rho}{d\varphi}\right|_{\varphi=0}\varphi + \left.\frac{d^2\rho}{d\varphi^2}\right|_{\varphi=0}\varphi^2 + \cdots\right] \tag{24}$$

with the charge density $\rho = e(n_i - n_{ep})$. Taking up to first order in Equation (24) and assuming quasi-neutrality at sheath edge, then multiplying with $\frac{d\varphi}{dx}$ and integrating with respect to $x$, it leads to:

$$E^2 + \frac{1}{\varepsilon_0}\left.\frac{d\rho}{d\varphi}\right|_{\varphi=0}\varphi^2 = C \tag{25}$$

where $C$ is a constant to be determined. Since electric field is 0 at sheath edge, clearly $C = 0$ and $\frac{d\rho}{d\varphi}\Big|_{\varphi=0}$ should satisfy:

$$\frac{d\rho}{d\varphi}\Big|_{\varphi=0} = -(\frac{E}{\varphi})^2 \leq 0 \tag{26}$$

Equation (26) can be further written as $\frac{d\rho}{d\varphi} = \frac{d\rho}{dx}\frac{dx}{d\varphi} = -E^{-1}\frac{d\rho}{dx} \leq 0$. It leads to the form of general Bohm criterion frequently used in literatures: $\frac{d\rho}{dx} \geq 0$ for ion sheath and $\frac{d\rho}{dx} \leq 0$ for electron sheath. Caution should be taken regarding the sign here.

With Equation (26), we can move on and derive the electron and ion density gradient at sheath edge. To do that, electron and ion fluid equations in presheath are given as follows:

$$\frac{d}{dx}(n_i u_i) = S_{ni} \tag{27}$$

$$\frac{d}{dx}(n_i T_i) = e n_i E + S_{mi} \tag{28}$$

$$\frac{d}{dx}\left(n_{ep} u_e\right) = S_{ne} \tag{29}$$

$$\frac{d}{dx}\left(m_e n_{ep} u_e^2 + n_{ep} T_{ep}\right) = -e n_{ep} E + S_{me} \tag{30}$$

where $u_i$ is ion fluid velocity, $S_{ni}$, $S_{ne}$, $S_{mi}$ and $S_{me}$ are source terms for electron and ion particle and momentum balance. Comparing Equation (28) and (30), one can find that the force on electrons contributed by electric field is roughly $\frac{T_i}{T_{ep}}$ times smaller than the pressure gradient term. Oppositely, it is the electric field that plays the dominant role in the ion-rich plasma sheath. Similar conclusions have been proposed by Brett et al.[17] At the sheath edge, we drop the source terms near sheath edge and obtain the ion and electron density gradients in the following form:

$$\frac{dn_i}{dx} = \frac{e n_i E}{T_i} \tag{31}$$

$$\frac{dn_{ep}}{dx} = \frac{e n_{ep} E}{m_e u_e^2 - T_{ep}} \tag{32}$$

Note that $n_{ep} = n_i$ at sheath edge. Using Equation (26), the Bohm criterion for electron sheath is derived as follows:

$$u_e|_{sheath\ edge} = u_{e0} \geq \sqrt{\frac{T_{ep}+T_i}{m_e}} \tag{33}$$

Equation (33) is analogous to Bohm criterion for ion sheath except that electron entering velocity is $\sqrt{\frac{m_i}{m_e}}$ times larger than ion entering velocity.

### 3.2.2 The role of kinetic ions on electron sheath Bohm criterion

In the deductions above, ions are assumed to be balanced by pressure gradient and electric field just as plasma electrons in an ion sheath. Actually, the EVDF in an ion sheath is commonly known as being depleted at high velocity end since energic electrons that penetrate the ion sheath will not return back to plasma. This phenomenon is known as the electron loss cone.[27] The effects of depleted EVDF at high velocity tail in ion sheath have been addressed in detail by numerous works.[28-30] Particularly, Joaquim et al showed that the Bohm velocity of an ion sheath could be smaller than the conventional ion sound velocity $\sqrt{\frac{T_{ep}+T_i}{m_i}}$ when the loss cone is considered.[31] Following a similar logic, it is intuitive to imagine that the Bohm velocity of the electron sheath $u_{e0}$ could also be affected by IVDF depleted at high velocity end, though the influence can only be obvious at high ion temperature.

Similar to loss cone in the ion sheath, fast ion may penetrate the potential barrier in electron sheath and thus will not return, leading to the following truncated velocity distribution function:

$$f_i = n_i(\varphi)\begin{cases}\frac{1}{I(\eta)}\sqrt{\frac{m_i}{2\pi T_i}}\exp\left(-\frac{m_i v_i^2}{2T_i}\right), -\infty \leq v_i \leq v_{icut} \\ 0, v_i \geq v_{icut}\end{cases} \tag{34}$$

Here normalized potential term is $\eta = \frac{e(\varphi_w - \varphi)}{T_i}$ with $\varphi_w$ the potential at electrode (to be distinguished from $\Phi$ which is normalized with respect to $T_{ep}$), ion cutoff velocity is $v_{icut} = \sqrt{\frac{2e(\varphi_w-\varphi)}{m_i}} = \sqrt{2\eta}v_{thi}$, electrode potential relative to sheath edge is $\varphi_w$, and $I(\eta) = 0.5[1 +$ $\mathrm{erf}(\sqrt{\eta})]$. Ion flow velocity is then calculated as $u_i = \frac{v_{thi}}{I(\eta)}\exp(\Lambda - \eta)$, where ion thermal velocity is $v_{thi} = \sqrt{\frac{T_i}{m_i}}$ and $\Lambda = -\ln(\sqrt{2\pi})$. Its derivation with respect to $\varphi$ is $\partial_\varphi u_i = \frac{eu_i}{T_i}(1 + \frac{\exp(-\eta)}{2\sqrt{\pi\eta}I(\eta)})$, to be used later on.

Now that we have the ion flux term, reusing Equation (27), (29), (30), and the three equations are restructured into the matrix form $MX = S$ with $X = (\partial_x n, \partial_x u_e, \partial_x \varphi)^T$ , $S = (S_{ni}, S_{ne}, S_{me})^T$ and matrix $M$ in the following form:

$$M = \begin{pmatrix} u_i & 0 & n\partial_\varphi u_i \\ u_e & n & 0 \\ T_{ep} & m_e n u_e & -en \end{pmatrix} \tag{35}$$

Note that in presheath the quasi-neutrality is satisfied so electron and ion densities are both $n$.

Since the gradient terms become much larger near sheath edge, we can set $S = 0$ and solve for $u_e$ using $\det(M) = 0$ in order to get a nontrivial set of solution. We arrive at:

$$u_e = \sqrt{\frac{T_{ep} + T_i/(1+\kappa)}{m_e}} = u_{the}\sqrt{1 + \frac{T_i}{T_{ep}}\frac{1}{1+\kappa}} \tag{36}$$

with $\kappa = \frac{\exp(-\eta_{se})}{2\sqrt{\pi\eta_{se}}I(\eta_{se})}$ (subscript $se$ means sheath edge so $\eta_{se} = \frac{e\varphi_w}{T_i}$) and electron thermal velocity $v_{the} = \sqrt{\frac{T_{ep}}{m_e}}$. The electron entering velocity can be written as $u_{e0} = \alpha_{ue0}u_{the}$, with the factor $\alpha_{ue0}$ in the following form:

$$\alpha_{ue0} = \sqrt{1 + \frac{1}{\Theta_T}\frac{1}{1+\kappa}} \tag{37}$$

Above equation suggests that the electron sheath can become super-thermal, i.e. the electron entering velocity at sheath is higher than the thermal velocity predicted by Bohm criterion. Calculation results are shown in Figure 5(a). Clearly, electron flow velocity at sheath edge is always above the electron thermal velocity $u_{the}$, and increases with $\eta_{SE}$. It finally saturates at $v_{e,max} = \sqrt{\frac{T_{ep}+T_i}{m_e}}$ when $\eta_{SE} \to \infty$. The influence of $\eta_{SE}$ is actually mitigated if $\Theta_T$ is large. Meanwhile, ion flow velocity is always below its thermal velocity and decreases with $\eta_{SE}$.

One may also estimate ion heat flux $q_i = \frac{1}{3} m_i n_i \langle (v_x - u_i)^3 \rangle$ by calculating $\langle v_x^2 \rangle$ and $\langle v_x^3 \rangle$ according to the distribution function in Equation (34). Ion heat flux is expressed as follows:

$$q_i = \frac{m_i n_i v_{thi}^3}{\sqrt{2\pi} I(\eta)} \left[\exp(-\eta)\left(\eta - \frac{1}{2}\right) + \frac{3}{2}\sqrt{\frac{\eta}{\pi}} \frac{\exp(-2\eta)}{I(\eta)} + \frac{\exp(-3\eta)}{2\pi I^2(\eta)}\right] \tag{38}$$

Note that both $v_i$ and $q_i$ are independent from $\Theta_T$. Calculation results are given in Figure 5(b)(c). It is reasonable that the ion flux decreases with $\eta_{SE}$ since a larger potential barrier obviously diminishes the ion flow in the electrons sheath, and the ion flux goes to 0 at infinite $\eta_{SE}$. The ion heat flux, however, first rises up and then declines as $\eta_{SE}$ increases. This trend is similar to the electron heat flux in a subsonic ion sheath and shows interesting symmetry between the two types of sheath.[31]

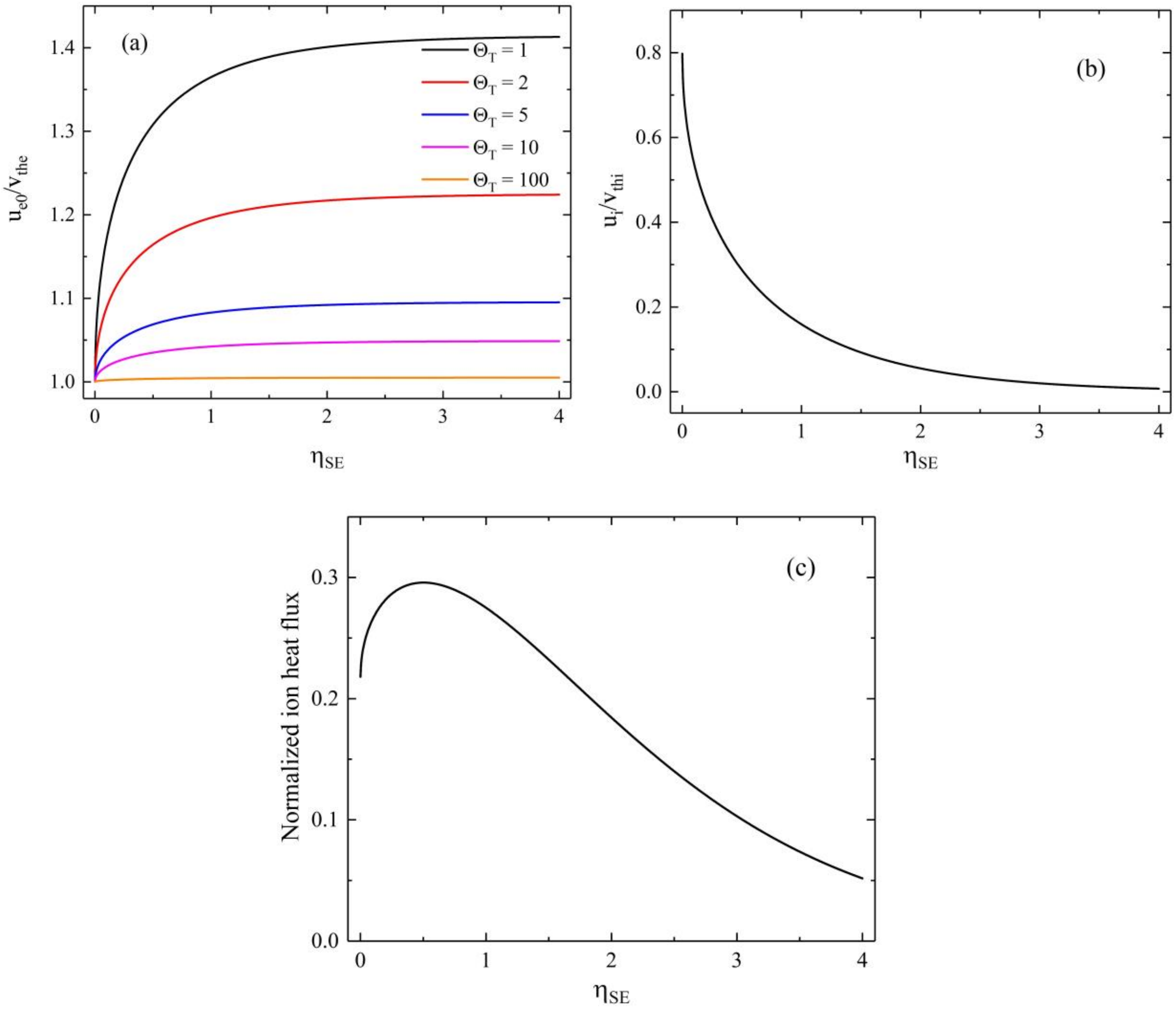

Figure 5. Sheath edge properties as function of term $\eta_{SE} = \frac{e\varphi_w}{T_i}$ and $\Theta_T = \frac{T_{ep}}{T_i}$. (a) electron entering velocity at sheath edge normalized with respect to electron thermal velocity. Electrons velocity increases

with $\eta_{SE}$ while its influence becomes smaller at higher $\Theta_T$. (b) Ion velocity normalized over ion thermal velocity, which decreases with $\eta_{SE}$. (c) ion heat flux normalized over $\frac{1}{2} m_i n_{SE} v_{th}^3$.

To summarize above, in the section various forms of electron entering velocity are introduced. When neglecting all ions, it is simply $u_{e0} = \sqrt{\frac{T_{ep}}{m_e}}$. If considering nonzero ion temperature but disregarding the ion loss cone, the expression becomes $u_{e0} = \sqrt{\frac{T_{ep}+T_i}{m_e}}$. If one further involves the ion loss cone, the expression is structured as $u_{e0} = \alpha_{ue0} u_{the}$ with $\alpha_{ue0} = \sqrt{1 + \frac{1}{\Theta_T}\frac{1}{1+\kappa}}$. These expressions will be used in derivation of the following sections.

### 3.3 Application of improved Bohm criterion in kinetic modeling of electron sheath

#### 3.3.1 The truncated EVDF and its influence on kinetic modeling of electron sheath

In section 3.2, we derive the electron entering velocity $u_{e0}$ in various types of assumption. It is then possible to apply the obtained $u_{e0}$ to revise the adopted form of EVDF at sheath entrance. Recall that in section 3.1, the EVDF at sheath edge is regarded as being in thermal equilibrium and has zero drift velocity. When $u_{e0}$ is included, however, the EVDF at sheath edge is inevitably truncated, and a drift velocity should be included.

For simplicity, we assume that the EVDF is simply a half-Maxwellian with no drift velocity, and then add up a nonzero $u_{e0}$ to see its influence. The comparison of the EVDFs with and without entering velocity is given in Figure 6.

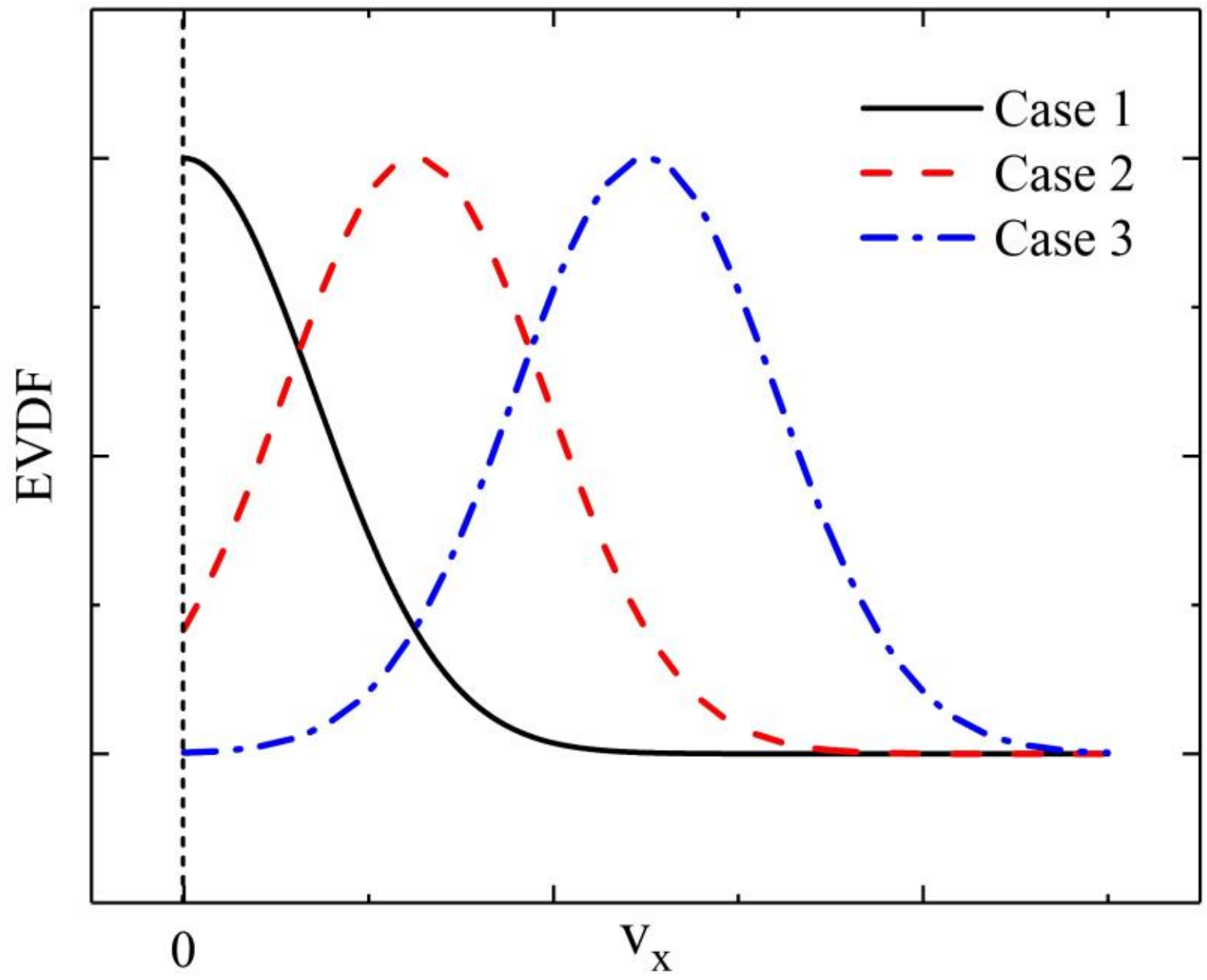

Figure 6. The schematic of the EVDFs with zero or nonzero entering velocity. EVDFs are normalized with respect to the density. Case 1 is the half-Maxwellian case, case 2 is an arbitrary nonzero entering velocity case, while case 3 is similar to 2 except that the drift velocity is large so that EVDF is close to a drifting Maxwellian.

The half-Maxwellian EVDF in Figure 6 is simply $f_{e,hM} = n_{se}\sqrt{\frac{2m_e}{\pi T_{ep}}}\exp(-\frac{m_e v_e^2}{2T_{ep}})$. The truncated EVDFs in Figure 6 are as follows:

$$f_{e,ue0} = n_{se}\frac{1}{1+\mathrm{erf}(u_{e0}\sqrt{\frac{m_e}{2T_{ep}}})}\sqrt{\frac{2m_e}{\pi T_{ep}}}\exp[-\frac{m_e(v_e-u_{e0})^2}{2T_{ep}}], v_e \geq 0 \tag{39}$$

The additional factor in Equation (39) with regard to half-Maxwellian is derived by normalization in terms of the sheath edge density $n_{se}$. Note that when $u_{e0} = 0$, Equation (39) is reduced to the half-Maxwellian used in section 3.1, whereas when $u_{e0}$ is larger relative to electron thermal velocity, the EVDF approaches a flowing Maxwellian. This suggests that the fluid model employed in section 2.2 is an approximation of infinite electron entering velocity.

The electron flux at sheath edge is then calculated as:

$$\Gamma_{ep,ue0} = \int_0^{+\infty} v_e f_{e,ue0} dv_e = \beta_e \Gamma_{ep,hM} \tag{40}$$

where $\Gamma_{ep,hM} = n_{se}\sqrt{\frac{2T_{ep}}{\pi m_e}}$ is the electron flux at sheath edge for half-Maxwellian EVDF, and $\beta_e$ is a coefficient depending on $u_{e0}$ in the following form:

$$\beta_e = \frac{\exp\left(-\frac{m_e u_{e0}^2}{2T_{ep}}\right)}{1+\text{erf}(u_{e0}\sqrt{\frac{m_e}{2T_{ep}}})} + u_{e0}\sqrt{\frac{\pi m_e}{2T_{ep}}} \tag{41}$$

Equation (41) can be justified by the limit $u_{e0} = 0$ which yields $\beta_e = 1$ and is reduced back to the half-Maxwellian case. For typical low temperature plasma with $T_{ep} \gg T_i$, $u_{e0} = \sqrt{\frac{T_{ep}}{m_e}}$ and $\beta_e \approx 1.61$. Recall the conclusion obtained in section 3.2 that $u_{e0} = \alpha_{ue0} u_{the}$ where $u_{the} = \sqrt{\frac{T_{ep}}{m_e}}$ and $\alpha_{ue0}$ depends on the electron-ion temperature ratio $\Theta_T$ and $\kappa(\Phi_w)$. The calculated results of factor $\beta_e$ are shown in Figure 7.

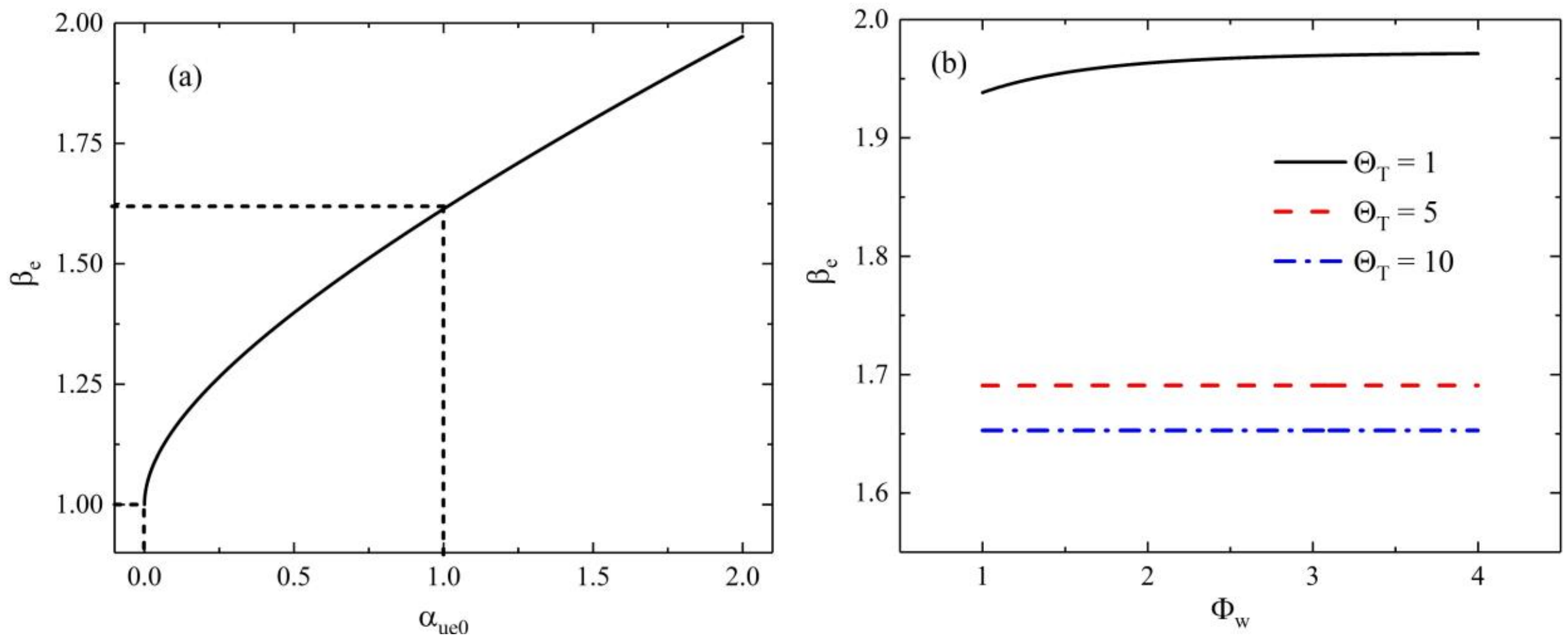

Figure 7. Calculated $\beta_e$ as a function of (a) $\alpha_{ue0}$ and (b) $\Phi_w$ and $\Theta_T$.

It is clear that the value of $\beta_e$ increases with $\alpha_{ue0}$. Note that the presheath-sheath matching requires that $\alpha_{ue0} \in [1,2]$, which dictates a minimum $\beta_e$ of around 1.61 and maximum of around 1.97. In Figure 7(b) one can find that the influence of $\Phi_w$ on $\beta_e$ is obvious at low $\Theta_T$ level, and the $\Theta_T$ plays more dominant role compared with $\Phi_w$. This is as expected because the ion loss cone is obvious only when sufficient amounts of ions can penetrate the electron sheath potential. At higher $\Theta_T$ level, the normalized electrode potential $\Phi_w$ is not influential. This is reassuring for low temperature plasma applications such as Langmuir probe, since if the entering velocity changes with electrode biased potential, the probe measurement must therefore be calibrated accordingly. However, for lower $\Theta_T$ ($\Theta_T < 1$), the influence of $\Phi_w$ is quite remarkable (not shown here), corresponding to for instance the plasma in divertor region of magnetic fusion device.[32] This might rise concern for probe data processing but is beyond the scope of the present work. Future experimental works in fusion background are expected to further address this issue.

### 3.3.2 An attempt to include Bohm criterion in the kinetic model of electron sheath.

In section 3.1, a kinetic model is established for electron sheath involving the proper EVDF at sheath edge, which provides numerical solution of the sheath structure. However, the electron entering velocity was not counted in that model. It is therefore intuitive to consider if it is possible to modify the form EVDF at sheath edge according to the revised electron Bohm criterion and apply similar methods to further improve the kinetic modeling. We will show in this section that it is not as straightforward as anticipated, and not even a numerical solution could be obtained when the entering velocity is considered.

In an arbitrary position of the electron sheath, its EVDF should contain the drift velocity $u_{e0}$ imposed by Bohm criterion and the electron cutoff velocity $v_{ecut}$ due to field acceleration. Note that $u_{e0} = \alpha_{ue0} u_{the}$ and $v_{ecut} = \sqrt{2\Phi_W \frac{T_{ep}}{m_e}}$. In order to apply the similar methods in section 3.1, the electron density should be evaluated first. The integral of the EVDF depends on the relative position between $u_{e0}$ and $v_{ecut}$ in the velocity axis, as shown in figure below.

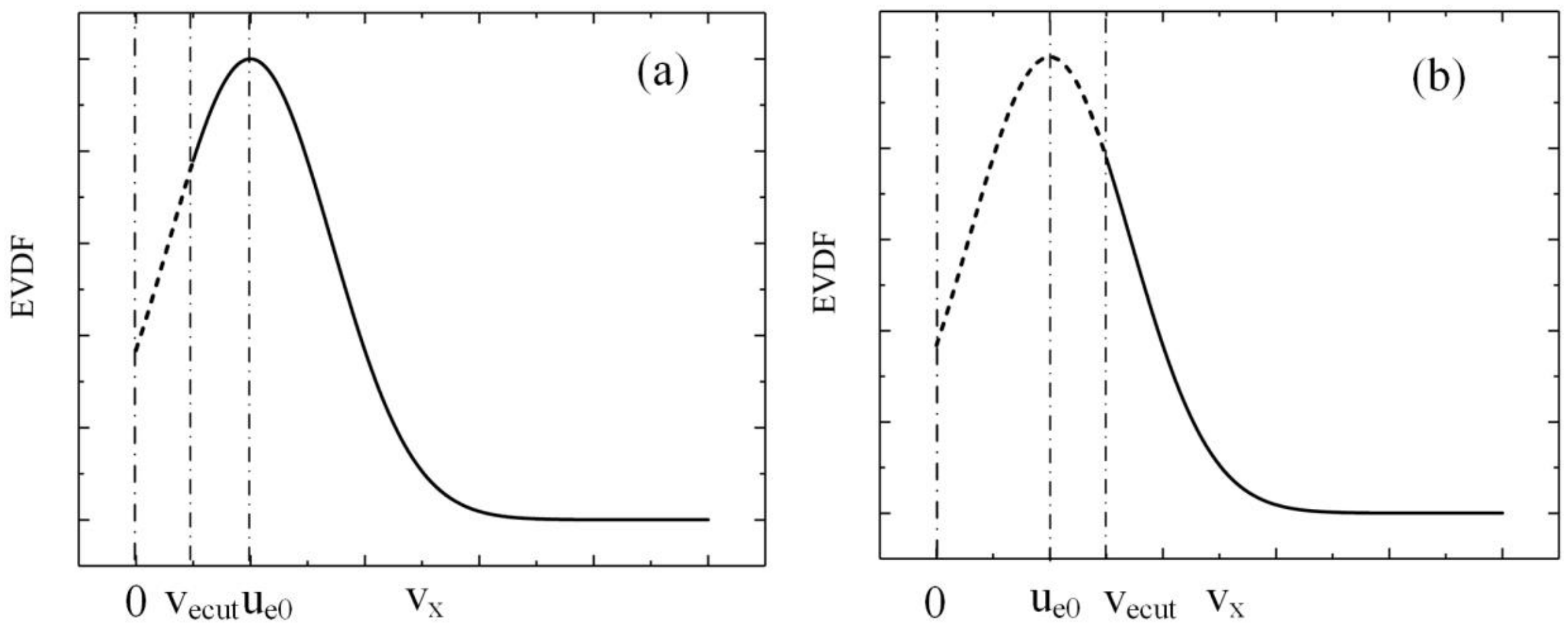

Figure 8. Schematic of two cases of EVDF and the influence of the relation between $u_{e0}$ and $v_{ecut}$.

In case 1 the electron density is calculated to be:

$$n_{ep1} = n_{se}\exp(\frac{e\varphi}{T_{ep}})\frac{1+\text{erf}[(u_{e0}-v_{ecut})\sqrt{\frac{m_e}{2T_{ep}}}]}{1+\text{erf}(u_{e0}\sqrt{\frac{m_e}{2T_{ep}}})} \tag{42}$$

When $v_{ecut} = 0$, $\varphi = 0$, which corresponds to the sheath edge, we get $n_{ep}\big|_{se} = n_{se}$. For case 2, the integrated electron density is in the following form:

$$n_{ep2} = n_{se}\exp(\frac{e\varphi}{T_{ep}})\frac{\text{erfc}[(v_{ecut}-u_{e0})\sqrt{\frac{m_e}{2T_{ep}}}]}{1+\text{erf}(u_{e0}\sqrt{\frac{m_e}{2T_{ep}}})} \tag{43}$$

Here when $u_{e0} = 0$, the expression is reduced to the Equation (18) where no entering velocity is considered. Therefore, the two expressions above can be justified by the limits. Also note that the above two equations are continuous at $v_{ecut} = u_{e0}$.

However, when bringing the electron density expression into the Poisson equation, no analytical expression can be obtained from the integral of Equation (42) and (43), so not even the numerical solution of electron sheath as in section 3.1 are derivable. This indicates that a compromise has to be made when choosing the electron sheath model: either the electron sheath Bohm criterion is counted and fluid model is employed (section 2.2) or the kinetic effects (EVDF) is considered while entering velocity must be dropped. Future works are expected to propose better solution and self-consistently combined the truncated EVDF and electron sheath solution.

# 4 Electron sheath modeling considering plasma-surface interaction

Secondary electron emission widely exists in a multitude of scenarios where plasma flux coming from sheath contacts the solid boundaries.[27, 29, 30, 33, 34] Both incident ion and electron can induce SEE at solid boundary. The former is generally through Auger neutralization or deexcitation where approaching ion bends the local field near solid material so that the trapped electrons can escape from the wall.[35] The latter happens when the incident primary electron is deaccelerated after penetrating the material surface while transferring energy to surrounding electrons, some of which could evacuate from material surface and becomes true secondary electrons.[36]

In low temperature plasma applications, both types of SEE could happen, but not both of them are considered simultaneously in most if not all plasma simulations. The ion-induced SEE coefficient is independent from the incident ion temperature (only true for cold ion) and is a constant once the wall material is fixed. Recent theories showed that the accumulated surface charges in dielectric materials can modify the ion-induced SEE coefficient,[37] but here we focus on metallic boundary. On the other hand, the electron-induced SEE usually depends on primary electron energy. The ion-induced SEE is usually counted if the electron temperature is low and the wall material is metallic which has lower electron-induced SEE coefficient. Taking the example of plasma processing using capacitively-coupled plasma, it is mostly the ion-induced SEE that is counted since electrodes are typically metallic and electron temperature is low,[38] whereas the electron-induced SEE was shown to be influential if the neutral pressure is low and the electrodes are made of more emissive material such as $SiO_2$.[39]

In the particular case of the electron sheath, the reason why we draw special attention to SEE is that the plasma electrons are accelerated by the sheath potential. The electron incident energy for an ion sheath is $2T_{ep}$, whereas the incident energy for an electron sheath is $2T_{ep} + e\varphi_w$. This marks the possible significance of electron-induced SEE (simply called SEE in the following section) even with metallic electrode.

To include SEE into electron sheath theory, it is important to first clarify the electron dynamics therein. The plasma electrons are accelerated, colliding on electrode and causing SEE. Emitted electrons are repelled by the sheath potential and some of them are returned back to the electrode. Since the reflected emitted electron energy is usually much smaller than the energy of sheath-accelerated plasma electrons, it is assumed that the reflected electrons no longer cause SEE. The following derivations are inspired by the highly emissive inverse sheath model.[40] Note that the inverse sheath is different from the electron sheath discussed here, since it has no presheath and a emission coefficient greater than 1 is required. The inverse sheath does not require strongly biased electrode and can appear on large, floating surface (no need for $A_w \ll A_c$). The electron fluxes at electrode for the electron sheath should satisfy the following equation:

$$\Gamma_{ep} + \Gamma_{eref} - \Gamma_{em} = \Gamma_{net} \tag{44}$$

where $\Gamma_{eref}$, $\Gamma_{em}$, and $\Gamma_{net}$ are reflected electron flux, emitted electron flux and the net electron flux, respectively. $\Gamma_{net}$ must be balanced with flux at other chamber surfaces where ion sheath presents.

The reflected electron flux is in the following form:

$$\Gamma_{eref} = \Gamma_{em}[1 - \exp\left(-\frac{e\varphi_w}{T_{em}}\right)] \tag{45}$$

Bringing in the definition of SEE coefficient $\gamma_e = \Gamma_{em}/\Gamma_{ep}$, the following equation can be derived:

$$\Gamma_{net} = \Gamma_{em}[\gamma_e^{-1} - \exp\left(-\frac{e\varphi_w}{T_{em}}\right)] \tag{46}$$

The expression of plasma electron flux at sheath edge is again $\Gamma_{ep} = n_{sep}\beta_e\sqrt{\frac{2T_{ep}}{\pi m_e}}$, note that here the plasma electron density at sheath edge $n_{sep}$ is not equal to the sheath edge density $n_{se}$ since $n_{se}$ contains both plasma electrons and emitted electrons. Regarding the distribution function of emitted electrons, the half-Maxwellian is frequently chosen to facilitate the calculation, though it was also shown that the choice of EVDF for emitted electrons can bring some influences in sheath solution.[41] The emitted electron flux at electrode is simply:

$$\Gamma_{em} = n_{emw}\sqrt{\frac{2T_{em}}{\pi m_e}} \tag{47}$$

with $n_{emw}$ the emitted electron density at electrode. At sheath edge, the total electron density is summed up to be $n_{se}$, which gives the following equation:

$$n_{emw}\exp\left(-\frac{e\varphi_w}{T_{em}}\right) + n_{sep} = n_{se} \tag{48}$$

Combining above equations, we arrive at the following expression for electron sheath potential:

$$\Gamma_{net} = n_{se}\sqrt{\frac{2T_{em}}{\pi m_e}}\frac{\gamma_e^{-1}-\exp\left(-\frac{e\varphi_w}{T_{em}}\right)}{\exp\left(-\frac{e\varphi_w}{T_{em}}\right)+\sqrt{\Theta_{Tem}}(\gamma_e\beta_e)^{-1}} \quad (49)$$

where $\Theta_{Tem} = T_{em}/T_{ep}$. It is interesting to find that the electron sheath potential can actually be derived directly here when SEE is considered. Remember that for either fluid model or kinetic model without SEE, the electron sheath potential $\varphi_w$ is always a given factor. The electron sheath potential is solved as:

$$\Phi_w = \Theta_{Tem}\ln\left(\gamma_e\frac{1+\chi}{1-\chi\sqrt{\Theta_{Tem}}\beta_e^{-1}}\right) \quad (50)$$

with $\Phi_w$ the normalized electrode potential, and the factor $\chi$ a normalized term in the following form:

$$\chi = \frac{\Gamma_{net}}{n_{se}\sqrt{\frac{2T_{em}}{\pi m_e}}} \quad (51)$$

The reason why considering SEE gives the solution of $\Phi_w$ but not if no SEE is involved is as follows: when no SEE is considered, the full plasma electron current is collected at electrode which is independent from $\Phi_w$, whereas when SEE is counted, the flux balance at electrode (Equation (44)) provides an additional equation linking $\Phi_w$ and $\Gamma_{ep}$, which makes $\Phi_w$ solvable. It has to be noted that the introduced factor $\chi$ is not straightforward in terms of physical meaning on its own, but the denominator term in Equation (50) dictates a singularity at $1 = \chi\sqrt{\Theta_{Tem}}\beta_e^{-1}$ which is simplified as $\Gamma_{ep} = \Gamma_{net}$. According to Equation (44) and (45), $\Gamma_{ep} - \Gamma_{em}\exp\left(-\frac{e\varphi_w}{T_{em}}\right) = \Gamma_{net}$. Therefore, the singularity is never achieved as long as secondary electron emission exists. While the upper bound of possible $\chi$ value is prescribed by this singularity, the lower bound of $\chi$ should guarantee that $\Phi_w \geq 0$. The calculation results given by Equation (50) are shown in Figure 9.

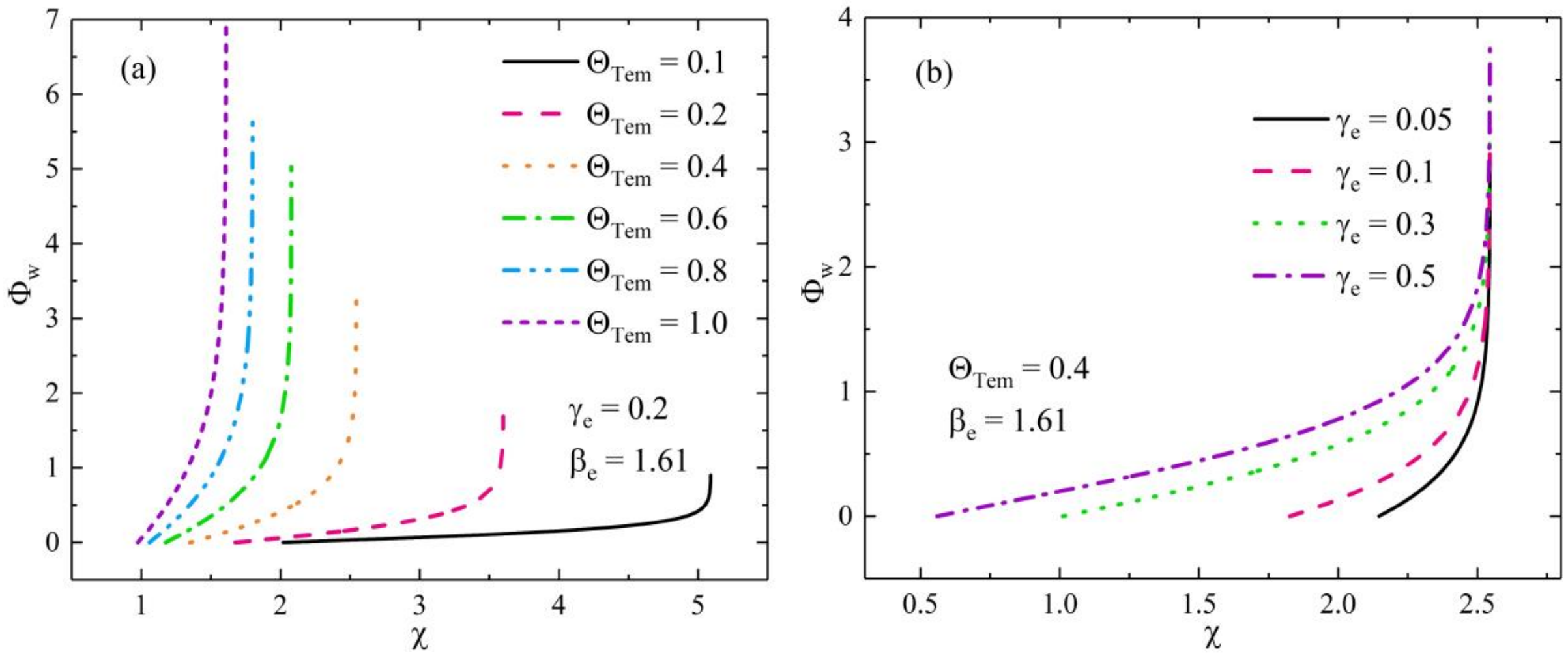

Figure 9. Calculated normalized electrode potential as function of $\chi$, $\gamma_e$, $\Theta_{Tem}$ and $\beta_e$.

It is obvious that the value of $\Theta_{Tem}$ has remarkable influence on the range of possible $\chi$. The value of $\beta_e$ varies between around 1.61 and 1.97, hence posing very limited influence on the profile of $\Phi_w$ (not shown). The emission coefficient mainly changes the lower bound of $\chi$ since a higher $\gamma_e$ allows for smaller $\chi$ value before the term $\gamma_e \frac{1+\chi}{1-\chi\sqrt{\Theta_{Tem}}\beta_e^{-1}}$ reaches 1.

However, the factor $\gamma_e$ is not an arbitrarily chosen factor. For thermionic emission or photoemission, the $\gamma_e$ can be regarded as constant once stable plasma sheath is formed,[42, 43] but here we focus on secondary electron emission where the emission coefficient is energy-dependent. The emission coefficient should be further expressed as:

$$\gamma_e = g(2T_{ep} + e\varphi_w) \tag{52}$$

where the form of $g(x)$ in the energy range of low temperature plasma can be approximated as $g(x) = x/\varepsilon_{see}$ or $g(x) = k\sqrt{x}$,[30] with $\varepsilon_{see}$ and $k$ material-dependent coefficients. For larger energy range, the SEE coefficient usually first increases with incident energy and then decreases after a certain threshold energy, such empirical model is widely employed in a variety of plasma simulations.[44, 45] For metallic materials, the value of $\varepsilon_{see}$ is usually around a decade eV.[46, 47] In practice, the situation is far more complex since the emission coefficient also depends on surface roughness, cleanness, and other local microstructures.[48, 49] For now, we simply give conceptual prediction based on the simplest form of emission coefficient: $\gamma_e = \frac{2T_{ep}+e\varphi_w}{\varepsilon_{see}} =$ $(2+\Phi_w)\Theta_{T\varepsilon}$, with $\Theta_{T\varepsilon}=\frac{T_{ep}}{\varepsilon_{see}}$. For cold plasma, the value of $\Theta_{T\varepsilon}$ should be smaller than 1, whereas in an attached divertor of fusion device $\Theta_{T\varepsilon}$ could reach or even surpasses 1. Here we focus on the typical cold plasma parameter condition.

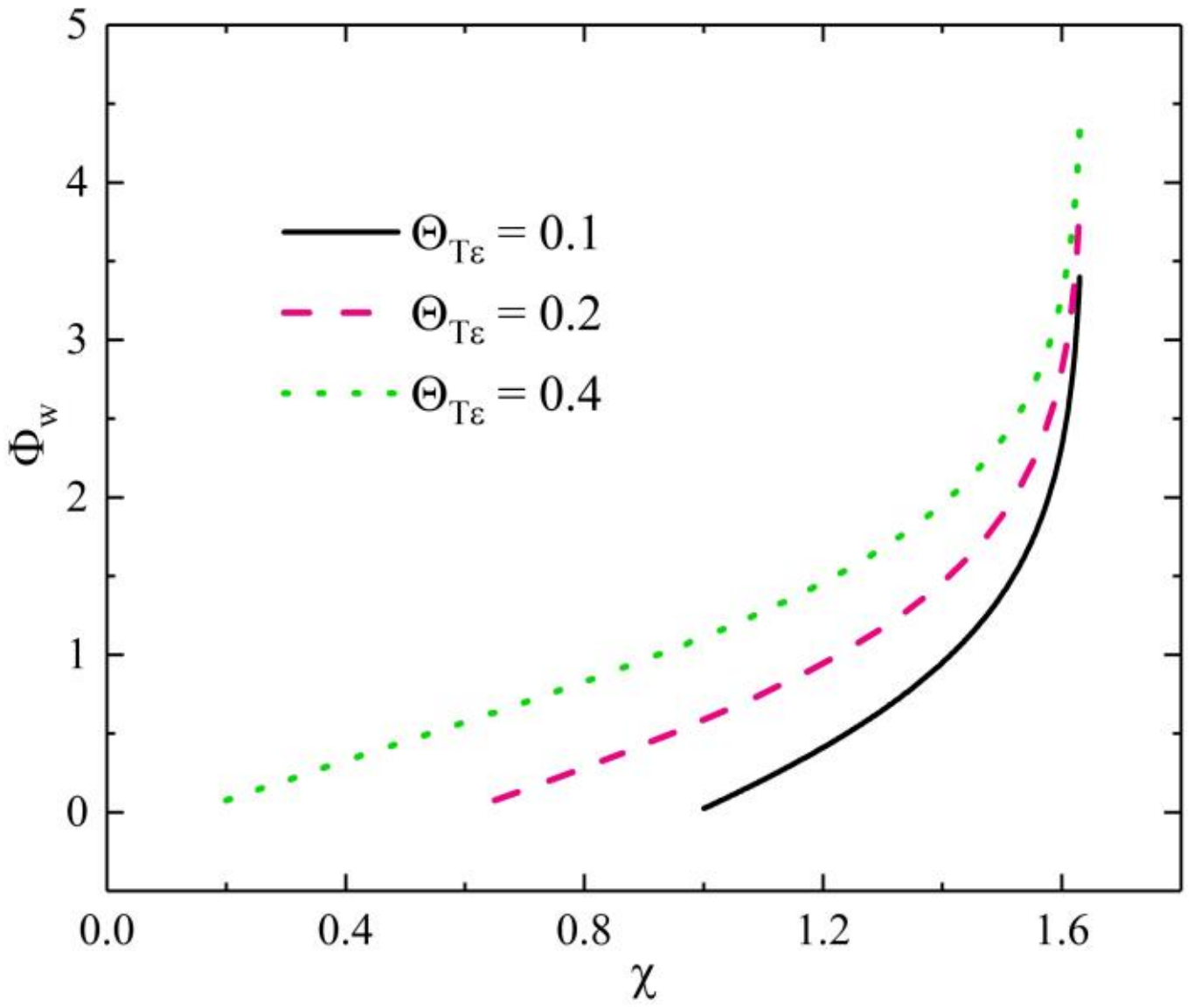

Figure 10. Calculated $\Phi_w$ considering the energy-dependent secondary electron emission coefficient.

In Figure 10 the lower bound of $\chi$ is shifted leftwards as $\Theta_{T\varepsilon}$ increases for the same reason in Figure 9(b). Now the $\gamma_e$ increases as $\Phi_w$ increases and the obtained profile is somewhat shifted compared with constant $\gamma_e$ case. Note that at low $\Phi_w$ level, the incident electrons might have a larger SEE coefficient due to reflection (also called backscattering), where the primary electron does not penetrate into the solid material but is elastically or inelastically returned. Assuming the reflection probability to be $R_f$, the expression of SEE coefficient has to be modified accordingly:

$$\gamma_{e,R}=R_f+(1-R_f)\gamma_e \tag{53}$$

where $\gamma_{e,R}$ is the electron emission coefficient involving electron reflection. This is because when a reflection occurs, the “effective” SEE coefficient is 1. However, the nature of such process is not as simple as we describe here and the coefficient $R_f$ is also energy dependent.[50] This requires more detailed quantum mechanical treatment and is beyond the interest of the present work.[30] In the end, the influence of SEE on electron sheath can be verified with experiment by using different types of electrodes (thus different $\gamma_e$) biased at the same electrode potential, then checking the electron sheath properties posed by different SEE coefficients of the electrodes. Future experimental works are expected to corroborate the theoretical predictions.

## 5 Conclusions

The present work is dedicated to an improved understanding of electron sheath theory and its implications in plasma-surface interaction. A fluid model considering the electron sheath entering velocity is proposed and compared to the classic electron sheath Child-Langmuir model. It is shown that the revised fluid model avoids the singularity at sheath edge in Child-Langmuir model and the latter tends to underestimate the electron sheath potential. Subsequently, a kinetic electron sheath model is constructed. Nonzero ion temperature is considered while disregarding the electron presheath structure. The kinetic model is shown to predict higher sheath potential, and the sheath size is reduced with higher ion temperatures. In the following, the electron presheath-sheath matching is developed extensively, utilizing a more realistic truncated ion distribution function due to the loss cone effect. The electron entering velocity at sheath edge is shown to be dependent on electron-ion temperature ratio and electron sheath potential, implying a potentially neglected calibration to make in probe diagnostic. Electron entering velocity is higher than Bohm criterion prediction when truncated IVDF is employed. Meanwhile, attempts are made to further include the electron entering velocity into kinetic models. The exact relation between density and potential is obtained while no analytic or numerical sheath potential solution can be derived, revealing a trade-off between electron presheath structure and kinetic treatment of electron sheath. In the end, the electron sheath theory is further generalized to include the secondary electron emission on the electrode induced by sheath-accelerated plasma electrons. It is found that the inclusion of surface electron emission provides an additional equation for electron sheath potential compared with the non-emissive kinetic models. In that case, analytical solution of electron sheath potential can be obtained instead of being given as prerequisite. Both constant and energy-dependent electron emission coefficient are employed in the derivation and the influence of backscattering is discussed. Further insights are also provided for future verifications of proposed theories.

## Acknowledgements

The research was conducted under the auspices of National Key R&D Program of China (Grant No. 2020YFC2201100) and the National Natural Science Foundation of China (GrantNo.52077169). This work was supported in part by the Swiss National Science Foundation.

## Appendix

The following table gives the key variables adopted in derivations.

| Notation | Meaning | Unit | Remarks |
|---|---|---|---|
| $\alpha_{ue0}$ | Normalized factor | 1 | $u_{e0} = \alpha_{ue0} v_{the}$ |
| $\beta_e$ | Normalized factor | 1 | $\Gamma_{ep,ue0} = \beta_e \Gamma_{ep,hM}$ |
| $\gamma_e$ | SEE coefficient | 1 | $\Gamma_{em}/\Gamma_{ep}$ |
| $\gamma_{e,R}$ | SEE coefficient | 1 | Consider reflection |
| $A_c$ | Area of chamber wall | $m^{-2}$ | Ion sheath presents |
| $A_w$ | Area of biased electrode | $m^{-2}$ | Electron sheath presents |
| $\Gamma_{ep}$ | Plasma electron flux | $m^{-2}s^{-1}$ | |
| $\Gamma_{ep,hM}$ | Plasma electron flux | $m^{-2}s^{-1}$ | Integrated with $f_{e,hM}$ |
| $\Gamma_{ep,ue0}$ | Plasma electron flux | $m^{-2}s^{-1}$ | Integrated with $f_{e,ue0}$ |
| $\Gamma_i$ | Ion flux | $m^{-2}s^{-1}$ | |
| $\Gamma_{net}$ | Net plasma flux | $m^{-2}s^{-1}$ | |
| $\Gamma_{eref}$ | Reflected electron flux | $m^{-2}s^{-1}$ | |
| $\Gamma_{em}$ | Emitted electron flux | $m^{-2}s^{-1}$ | |
| $\varepsilon_0$ | Vacuum permittivity | $CV^{-1}m^{-1}$ | |
| $\varepsilon_{see}$ | SEE model factor | $eV$ | |
| $f_{e,hM}$ | Half-Maxwellian EVDF | $m^{-4}s$ | |
| $f_{e,ue0}$ | Truncated EVDF | $m^{-4}s$ | |
| $f_i$ | IVDF | $m^{-4}s$ | |
| $E$ | Electric field | $Vm^{-1}$ | |
| $\eta$ | Normalized potential | 1 | $e(\varphi_w - \varphi)/T_i$ |
| $\eta_{se}$ | Normalized potential | | $\eta$ at sheath edge, $e\varphi_w/T_i$ |
| $\Theta_T$ | Temperature ratio | 1 | $T_{ep}/T_i$ |
| $\Theta_{Tem}$ | Temperature ratio | 1 | $T_{em}/T_{ep}$ |
| $\Theta_{T\varepsilon}$ | Temperature-energy ratio | 1 | $T_{ep}/\varepsilon_{see}$ |
| $I(\eta)$ | Normalized factor | 1 | $I(\eta) = 0.5[1 + \mathrm{erf}(\sqrt{\eta})]$ |
| $\kappa$ | Normalized factor | 1 | $\exp(-\eta_{se})\,/2\sqrt{\pi\eta_{se}}I(\eta_{se})$ |
| $\Lambda$ | Constant | 1 | $-\ln(\sqrt{2\pi})$ |
| $\lambda_{De}$ | Electron Debye length | $m$ | |
| $\mu$ | Ion-electron mass ratio | 1 | |
| $J_{net}$ | Net current density | $A$ | $e\Gamma_{net}$ |
| $m_e$ | Electron mass | $kg$ | |
| $m_i$ | Ion mass | $kg$ | |
| $n_{ep}$ | Plasma electron density | $m^{-3}$ | |
| $n_{sep}$ | Plasma electron density | $m^{-3}$ | Density at sheath edge |
| $n_{em}$ | Emitted electron density | $m^{-3}$ | |
| $n_{emw}$ | Emitted electron density | $m^{-3}$ | Density at biased electrode |
| $n_{se}$ | Density at sheath edge | $m^{-3}$ | |
| $n_i$ | Ion density | $m^{-3}$ | |
| $N$ | Normalized density | 1 | $n/n_{se}$ |
| $P_{e\parallel}$ | Electron parallel pressure | $Pa$ | $n_eT_{ep}$ |
| $q_i$ | Ion heat flux | $Jm^{-2}s^{-1}$ | |
| $R_f$ | Reflection coefficient | 1 | |
| $S_{ne}$ | Electron particle source | $m^{-3}s^{-1}$ | |
| $S_{ni}$ | Ion particle source | $m^{-3}s^{-1}$ | |
| $S_{me}$ | Electron momentum source | $kgm^{-2}s^{-2}$ | |
| $S_{mi}$ | Ion momentum source | $kgm^{-2}s^{-2}$ | |
| $T_{ep}$ | Plasma electron temperature | $eV$ | |
| $T_{em}$ | Emitted electron temperature | $eV$ | |
| $T_i$ | Ion temperature | $eV$ | |
| $u_{eo}$ | Electron entering velocity | $u_{eo}$ | Velocity at sheath edge |
| $u_e$ | Electron fluid velocity | $ms^{-1}$ | |
| $u_i$ | Ion fluid velocity | $ms^{-1}$ | |
| $v_{icut}$ | Ion cutoff velocity | $ms^{-1}$ | $\sqrt{2\eta}v_{thi}$ |
| $v_{ecut}$ | Electron cutoff velocity | $ms^{-1}$ | $\sqrt{2\Phi_w}v_{the}$ |
| $v_{the}$ | Electron thermal velocity | $ms^{-1}$ | $\sqrt{T_{ep}/m_e}$ |
| $v_{thi}$ | Ion thermal velocity | $ms^{-1}$ | $\sqrt{T_i/m_i}$ |
| $\varphi_{esh}$ | Electron sheath potential | $V$ | Positive |
| $\varphi_{ish}$ | Ion sheath potential | $V$ | Negative |
| $\varphi_w$ | Biased electrode potential | $V$ | Positive |
| $\Phi$ | Normalized potential | 1 | $e\varphi/T_{ep}$ |
| $\Phi_w$ | Normalized potential | 1 | $e\varphi_w/T_{ep}$ |
| $\chi$ | Normalized factor | 1 | $\Gamma_{net}/(n_{se}\sqrt{2T_{em}/\pi m_e})$ |
| $x_{ish}$ | Ion sheath size | $m$ | |
| $x_{esh}$ | Electron sheath size | $m$ | |
| $X$ | Normalized position | 1 | $x/\lambda_{De}$ |